# First-Principles Investigation of Auxetic Piezoelectric Effect in Nitride Perovskites

Yanting Peng[1], Zunyi Deng[1], Siyu Song[2], Gang Tang[2,*], Jiawang Hong[1, #]

[1]School of Aerospace Engineering, Beijing Institute of Technology, Beijing, 100081, China

[2]Advanced Research Institute of Multidisciplinary Science, Beijing Institute of Technology, Beijing 100081, China

E-mails:

*Gang Tang: gtang@bit.edu.cn;

#Jiawang Hong: hongjw@bit.edu.cn;

## Abstract

The recently reported auxetic piezoelectric effect, which acts as the electrical counterpart of the negative Poisson's ratio, is of significant technical importance for applications in acoustic wave devices. However, this electric auxetic effect has not yet been reported in perovskite systems. In this work, we employ first-principles calculations to investigate the piezoelectric properties of six polar nitride perovskites with the chemical formula $ABN_3$ (A = La, Sc, Y; B = W, Mo). Among these, all compounds except $ScMoN_3$ exhibit the auxetic piezoelectric effect, which is characterized by an unusually positive transverse piezoelectric coefficient, along with a positive longitudinal piezoelectric coefficient. This behavior is in sharp contrast to previously reported results in $HfO_2$, where both the longitudinal and transverse piezoelectric coefficients are negative. These unusual positive transverse piezoelectric coefficients originate from the domination of the positive internal-strain contribution. We further confirm the auxetic piezoelectric effect with finite electric field calculations. Our research enriches the understanding of the piezoelectric properties of nitride perovskites and provides a new compositional space for the design of novel auxetic piezoelectric materials.

# I. Introduction

Piezoelectric materials, which convert electrical energy into mechanical energy and vice versa, have extensive industrial and commercial applications in telecommunications,[1,2] medical imaging,[3] and ultrasonic devices.[4,5] Since Jaffe et al. discovered lead-zirconate-titanate [Pb(Zr,Ti)O$_3$, PZT] piezoelectric ceramics in 1954,[6] perovskite piezoelectric materials have attracted considerable attention because of their outstanding piezoelectric properties. In 2017, Xiong et al. experimentally synthesized an organic-inorganic perovskite [trimethylchloromethyl ammonium trichloromanganese(II), Me$_3$NCH$_2$ClMnCl$_3$, (TMCMMnCl$_3$)] with a large $d_{33}$ of 185 pC/N,[7] thereby paving the way for high-performance organic-inorganic perovskite piezoelectrics.[8] In 2021, Zakutayev et al. synthesized the polar LaWN$_3$ nitride perovskite, demonstrating a strong piezoelectric response (40 pm/V).[9] The successful synthesis of nitride perovskites not only opens new compositional space for the design of high-performance piezoelectric materials but also provides potential integration advantages over oxide perovskites in commercial nitride semiconductor devices.

Beyond the exploration of perovskite compounds with large piezoelectric coefficients, the discovery of new piezoelectric effects has also drawn significant attention over the past few years. Typically, the longitudinal ($d_{33}$) and transverse ($d_{31}$, $d_{32}$) piezoelectric coefficients of a material have opposite signs. However, Liu et al. recently predicted that in HfO$_2$, the longitudinal and transverse coefficients share the same sign, enabling a uniform expansion or contraction in all directions under an external electric field.[10] They termed this phenomenon the "electric auxetic effect", acting as the electrical analogue of a negative Poisson's ratio. Subsequently, Yang et al. experimentally verified this auxetic piezoelectric effect, also termed the type I auxetic piezoelectric effect, in Au/Nb:SrTiO$_3$ heterostructures.[11] Furthermore, they reported a type II auxetic piezoelectric effect, characterized by opposite signs of the transverse coefficients ($d_{31}$ and $d_{32}$) along two orthogonal in-plane directions. Under an external electric field, this type II effect (i.e., $d_{31} > 0$, $d_{32} < 0$, $d_{33} > 0$) leads to expansion along the longitudinal direction and one in-plane direction while causing contraction along the other transverse direction. However, intrinsically auxetic piezoelectric material systems reported to date remain very limited. Therefore, it is of great significance to explore whether the newly reported nitride perovskite systems exhibit auxetic piezoelectric effect.

In this work, we investigate the piezoelectric properties of six nitride perovskites with the chemical formula $ABN_3$ (A = La, Sc, Y; B = W, Mo). Using first-principles calculations, we find that $LaWN_3$ and $LaMoN_3$ exhibit the type I auxetic piezoelectric effect ($d_{31}$, $d_{32}$ and $d_{33} > 0$), $ScWN_3$, $YMoN_3$ and $YWN_3$ exhibit the type II auxetic piezoelectric effect ($d_{31} < 0$, $d_{32} > 0$, $d_{33} > 0$), and $ScMoN_3$ exhibits a conventional piezoelectric response. The auxetic piezoelectric effect in nitride perovskite is in sharp contrast to previously reported results in $HfO_2$, where both the longitudinal and transverse piezoelectric coefficients are negative. These unusual positive transverse piezoelectric coefficients originate from the domination of the positive internal-strain contribution. Finally, we confirm the auxetic piezoelectric effect with finite electric field calculations. These results significantly enrich our understanding of the piezoelectric behavior of nitride perovskites and highlight their potential for applications requiring precise control of mechanical deformations.

## II. Methods

All the calculations were performed using the Vienna Ab initio Simulation Package (VASP) based on the density functional theory (DFT).[12, 13] Projector augmented wave (PAW)[14] method was used and and PBEsol exchange-correlation functional[15] was used. The valence electronic configurations for the pseudopotentials were as follows: $3p^64s^23d^1$ for Sc, $4s^24p^65s^24d^1$ for Y, $6s^25d^1$ for La, $4p^65s^14d^5$ for Mo, $5p^66s^25d^4$ for W and $2s^22p^3$ for N. The plane-wave cut-off energy and the $k$-point meshes were determined individually for each nitride perovskite considered in Table SI in

Supplemental Material. The crystal structures were optimized until the energy convergence was less than $10^{-8}$ eV and the forces on the atoms were smaller than $10^{-3}$ eV/Å. The phonon dispersions were computed with the PHONOPY code.[16] The elastic constants $C_{ij}$ were calculated using the stress-strain methodology.[17] The piezoelectric stress coefficients $e_{ij}$ were calculated using density functional perturbation theory.[18, 19] The piezoelectric strain coefficients $d_{ij}$ are obtained from $d_{ij} = e_{ik}S_{kj}$,[20] where $S_{kj}$ represents the elastic compliance constants ($S_{kj} = C_{kj}^{-1}$). The structures were visualized using VESTA software.[21] The Bilbao Crystallographic Server (BCS) was used for group theory analyses.[22, 23]

## III. Results and discussion

### A. Structure Properties and Dynamic Stability

We selected six polar nitride perovskites with the chemical formula $ABN_3$ ($A^{3+}$ = Sc, Y, La; $B^{6+}$ = Mo, W) to investigate their piezoelectric properties. Among them, $LaWN_3$ has been experimentally synthesized, while the remaining five compounds are theoretically predicted.[9, 24] Among these structures, $ScMoN_3$ adopts an orthorhombic phase with the $Pna2_1$ (No. 33) space group, as shown in Fig. 1(a). $ScWN_3$ and $YBN_3$ (B = Mo or W) shows an orthorhombic phase with the $Pmn2_1$ (No. 31) space group, as depicted in Fig. 1(b). $LaBN_3$ (B = Mo or W) exhibits a trigonal phase with the $R3c$ (No. 161) space group, as illustrated in Fig. 1(c).

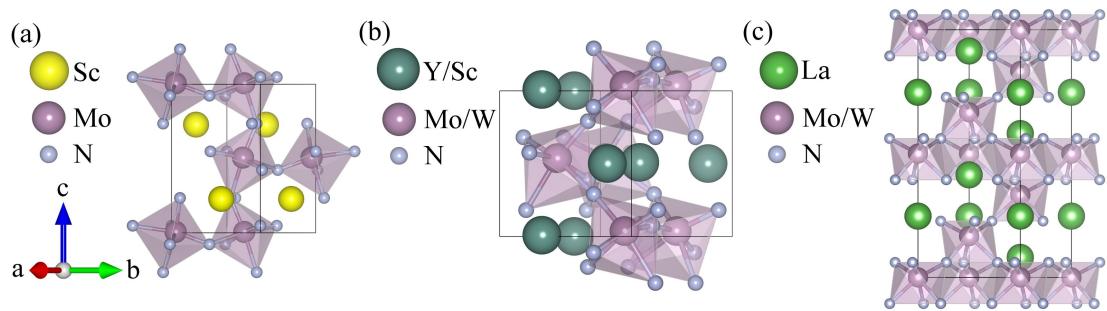

**FIG 1.** (a) Crystal structures of $ScMoN_3$ in the $Pna2_1$ phase. (b) $ScWN_3$ and $YBN_3$ (B = Mo or W) in the $Pmn2_1$ phase. (c) $LaBN_3$ (where B = Mo or W) in the $R3c$ phase.

The optimized lattice parameters for six nitride perovskites using the PBEsol functional are listed in Table I. For LaWN$_3$, the calculated lattice constants are $a$ = 5.651 Å, $b$ = 5.651 Å, $c$ = 13.724 Å, which are in good agreement with the experimental values ($a$ = 5.67 Å, $b$ = 5.67 Å, $c$ = 13.79 Å).[9] In addition, the predicted lattice parameters for the other five compounds are also consistent with previously reported theoretical results. As shown in Fig. 2(a-f), the calculated phonon spectra for the ABN$_3$ compounds exhibit no imaginary modes, confirming the dynamical stability of these polar phases. Furthermore, we evaluated the ferroelectric polarization of these compounds using the Berry phase method,[25] as shown in Table I. The results indicate that all six nitride perovskites exhibit polarization along the [001] direction. The calculated polarization values for LaMoN$_3$, YMoN$_3$, LaWN$_3$, ScMoN$_3$, ScWN$_3$, and YWN$_3$ are 68.738, 61.672, 60.861, 40.225, 35.095, and 27.496 μC/cm$^2$, respectively. Among them, the first three compounds exhibit polarization values exceeding 60 μC/cm$^2$, which is comparable to well-known ferroelectrics: such as LiNbO$_3$ (71 μC/cm$^2$)[26] and PbTiO$_3$ (75 μC/cm$^2$).[27]

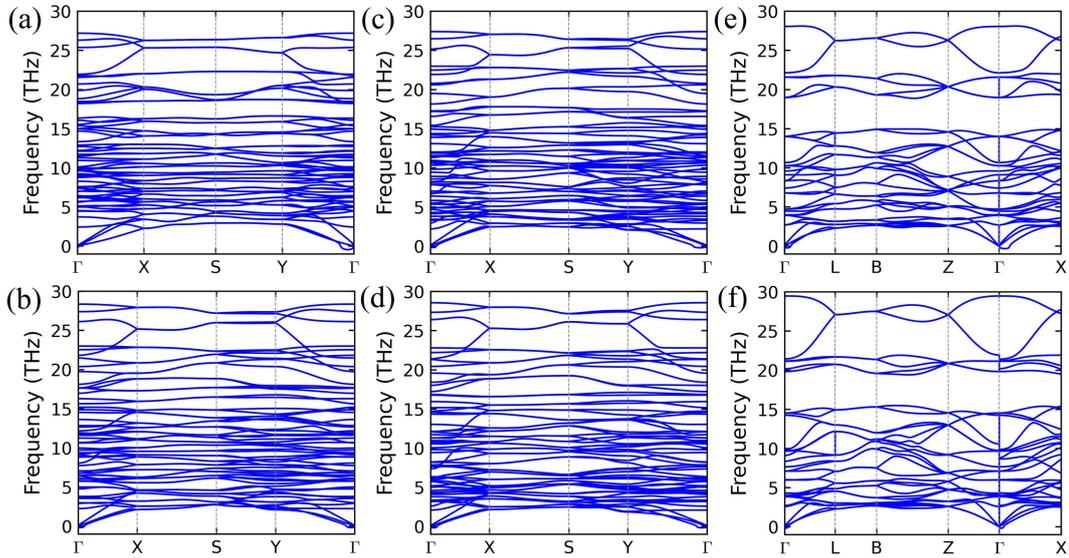

**FIG 2.** Phonon dispersion curves for the nitride perovskites (a) ScMoN$_3$, (b) ScWN$_3$, (c) YMoN$_3$, (d) YWN$_3$, (e) LaMoN$_3$ and (f) LaWN$_3$.

**TABLE I.** The optimized structural parameters of ABN$_3$ (A = Sc, Y, La; B = Mo, W) including space group (SPG), lattice constants a/b/c, polarization $P$.

| Compound | | SPG | $a$ [Å] | $b$ [Å] | $c$ [Å] | $P$ [μC/cm$^2$] |
|---|---|---|---|---|---|---|
| ScMoN$_3$ | this work | $Pna2_1$ | 5.552 | 5.158 | 7.570 | 40.225 |
| | Ref.[24] | $Pna2_1$ | 5.56 | 5.16 | 7.54 | 39 |
| ScWN$_3$ | this work | $Pmn2_1$ | 7.473 | 5.572 | 5.231 | 35.095 |
| | Ref.[24] | $Pmn2_1$ | 7.57 | 5.54 | 5.21 | 28 |
| YMoN$_3$ | this work | $Pmn2_1$ | 7.558 | 5.716 | 5.504 | 61.672 |
| | Ref.[24] | $Pmn2_1$ | 7.59 | 5.68 | 5.50 | 68 |
| YWN$_3$ | this work | $Pmn2_1$ | 7.649 | 5.663 | 5.470 | 27.496 |
| | Ref.[24] | $Pmn2_1$ | 7.74 | 5.63 | 5.45 | 34 |
| LaMoN$_3$ | this work | $R3c$ | 5.651 | 5.651 | 13.769 | 68.738 |
| | Ref.[28] | $R3c$ | 5.672 | 5.672 | 13.775 | 80.3 |
| LaWN$_3$ | this work | $R3c$ | 5.651 | 5.651 | 13.724 | 60.861 |
| | Ref.[29] | $R3c$ | 5.622 | 5.622 | 13.699 | 61 |
| | Ref.[9] | $R3c$ | 5.67 | 5.67 | 13.79 | - |

### B. Piezoelectric Coefficients

Based on the symmetries of the six nitride perovskites, ScMoN$_3$, ScWN$_3$, YMoN$_3$, and YWN$_3$ belong to the $mm2$ point group, while LaMoN$_3$ and LaWN$_3$ belong to the $3m$ point group. Therefore, the first four materials have five independent piezoelectric coefficients: $e_{31}$, $e_{32}$, $e_{33}$, $e_{24}$, and $e_{15}$. The latter two materials also have five independent coefficients: $e_{31}$, $e_{32}$, $e_{33}$, $e_{22}$, and $e_{15}$. The calculated piezoelectric stress coefficients ($e_{ij}$) are summarized in Table II. For further analysis, the total $e_{ij}$ is decomposed into two components:[30, 31] the clamped-ion contribution ($e_{ij}^{(0)}$) and the internal strain contribution ($e_{ij}^{(i)}$), both of which are also listed in Table II. Notably, among the six nitride perovskites, ScMoN$_3$ exhibits the largest longitudinal piezoelectric coefficient, with $e_{33}$ reaching 6.02 C/m$^2$. In addition, LaWN$_3$ shows the highest shear piezoelectric coefficient ($e_{15}$), with a value of 6.08 C/m², while LaMoN$_3$ also exhibits a comparably large $e_{15}$ of 5.49 C/m². Furthermore, the $e_{33}$ of all six compounds are found to exhibit same signs ($e_{33} > 0$), which can be primarily attributed to the dominant positive internal strain contribution ($e_{33}^{(i)} > 0$) in these materials. Interestingly, ScMoN$_3$ is the only compound that shows negative transverse piezoelectric coefficients ($e_{31}$ and $e_{32}$), which is consistent with the typical behavior observed in PbTiO$_3$,[32] where the longitudinal coefficient is positive and the transverse coefficients are negative. In contrast, the other five compounds display anomalously positive $e_{31}$ and $e_{32}$, except for ScWN$_3$, which has a negative $e_{31}$. Such unusual positive transverse piezoelectricity has also been recently reported in layered perovskites such as Ca$_3$Ti$_2$O$_7$ and Li$_2$SrNb$_2$O$_7$.[33]

Furthermore, based on the calculated elastic constants (see Supporting Information, Table SII) and piezoelectric stress coefficients ($e_{ij}$), we obtained the piezoelectric strain coefficients through the formula $d_{ij} = e_{ik}S_{kj}$, where $S_{kj}$ (see Supporting Information, Table SIII) represents the elastic compliance constants ($S_{kj} = C_{kj}^{-1}$), as shown in Table III. Notably, ScMoN$_3$, LaWN$_3$, and LaMoN$_3$ exhibit large longitudinal piezoelectric coefficients. In particular, ScMoN$_3$ shows the highest $d_{33}$ value, reaching 59.43 pC/N, significantly higher than that of other known nitrides, such as AlN (5.44 pC/N),[34] GaN (3.40 pC/N)[35] and SbN (20.27 pC/N).[36] The strong piezoelectric response mainly comes from the large $e_{33}$ and the large $S_{33}$ (see Supporting Information, Table SIII), according to $d_{33} = e_{31}S_{13} + e_{32}S_{23} + e_{33}S_{33}$. Interestingly, LaWN$_3$ and LaMoN$_3$ exhibit negative $d_{22}$ values, calculated to be -51.63 and -55.88 pC/N, respectively. In addition, LaWN$_3$, LaMoN$_3$, ScWN$_3$, YMoN$_3$, and YWN$_3$ exhibit high shear piezoelectric coefficients. Among them, LaMoN$_3$ exhibits the largest $d_{15}$ coefficient, reaching 180.80 pC/N, while LaWN$_3$ also shows a notably high $d_{15}$ value of 160.74 pC/N, higher than most inorganic oxide perovskites like PbTiO$_3$ (56.1 pC/N)[37] and LaNbO$_3$ (68 pC/N).[38] According to the relationship $d_{15} = e_{15}S_{55} + e_{16}S_{65}$ (where $e_{16} = -e_{22}$; $S_{44} = S_{55}$), the notably high $d_{15}$ likely arise from the large $e_{15}$ and large $S_{55}$ (see Supporting Information, Table SIII). Meanwhile, ScWN$_3$, YMoN$_3$, and YWN$_3$ show pronounced $d_{24}$ coefficients of 59.57, 55.10, and 57.06 pC/N, respectively.

**TABLE II.** Clamped-ion $e^{(o)}$ (C/m$^2$), internal strain $e^{(i)}$ (C/m$^2$), and total $e$ (C/m$^2$) piezoelectric stress constants of ABN$_3$ (A = Sc, Y, La; B = Mo, W).

|  |  | $e_{31}$ | $e_{32}$ | $e_{33}$ | $e_{15}$ | $e_{22}$ | $e_{24}$ |
|---|---|---|---|---|---|---|---|
| ScMoN$_3$ | $e^{(o)}$ | -0.07 | 0.59 | -0.29 | -0.10 | - | 0.33 |
|  | $e^{(i)}$ | -3.80 | -2.15 | 6.31 | -0.05 | - | 1.77 |
|  | $e$ | -3.87 | -1.56 | 6.02 | -0.15 | - | 2.10 |
| ScWN$_3$ | $e^{(o)}$ | 0.44 | -0.21 | 0.44 | 0.19 | - | -0.27 |
|  | $e^{(i)}$ | -0.74 | 3.42 | 2.95 | 0.51 | - | 4.57 |
|  | $e$ | -0.30 | 3.21 | 3.39 | 0.70 | - | 4.30 |
| YMoN$_3$ | $e^{(o)}$ | 0.82 | -0.21 | 0.55 | 0.52 | - | -0.69 |
|  | $e^{(i)}$ | -0.16 | 2.55 | 2.20 | 2.18 | - | 3.37 |
|  | $e$ | 0.64 | 2.29 | 2.75 | 2.80 | - | 2.68 |
| YWN$_3$ | $e^{(o)}$ | 0.54 | -0.26 | 0.44 | 0.22 | - | -0.39 |
|  | $e^{(i)}$ | -0.43 | 3.42 | 3.25 | 0.52 | - | 4.46 |
|  | $e$ | 0.11 | 3.16 | 3.69 | 0.74 | - | 4.07 |
| LaMoN$_3$ | $e^{(o)}$ | 0.19 | 0.19 | 0.62 | -0.40 | 0.70 | - |
|  | $e^{(i)}$ | 1.87 | 1.87 | 3.63 | 5.89 | -1.18 | - |
|  | $e$ | 2.06 | 2.06 | 4.25 | 5.49 | -0.48 | - |
| LaWN$_3$ | $e^{(o)}$ | 0.11 | 0.11 | 0.47 | -0.27 | 0.39 | - |
|  | $e^{(i)}$ | 2.13 | 2.13 | 4.11 | 6.35 | -2.18 | - |
|  | $e$ | 2.24 | 2.24 | 4.58 | 6.08 | -1.79 | - |

**TABLE III.** Calculated piezoelectric strain constants $d_{ij}$ (pC/N) of ABN$_3$ (A = Sc, Y, La; B = Mo, W).

|          | ScMoN$_3$ | ScWN$_3$ | YMoN$_3$ | YWN$_3$ | LaMoN$_3$ | LaWN$_3$ |
|----------|-----------|----------|----------|---------|-----------|----------|
| $d_{31}$ | -46.09    | -5.68    | -3.83    | -5.93   | 1.08      | 1.12     |
| $d_{32}$ | -14.86    | 6.25     | 7.16     | 7.89    | 1.08      | 1.12     |
| $d_{33}$ | 59.43     | 10.02    | 9.47     | 11.24   | 16.43     | 14.06    |
| $d_{15}$ | -1.12     | 5.99     | 25.93    | 5.91    | 180.80    | 160.74   |
| $d_{22}$ | -         | -        | -        | -       | -51.63    | -55.88   |
| $d_{24}$ | 22.16     | 59.57    | 55.10    | 64.33   | -         | -        |

## C. Auxetic Piezoelectric Effect

**Type I Auxetic Piezoelectric Effect.** Based on the above discussion and Table II and III, LaBN$_3$ (B = Mo, W) exhibits unusual positive transverse piezoelectric coefficients (i.e., $e_{31} > 0$ and $e_{32} > 0$). Combined with the positive longitudinal piezoelectric coefficients ($e_{33} > 0$), this indicates that LaBN$_3$ (B = Mo, W) possesses a type I auxetic piezoelectric effect. To further confirm the auxetic piezoelectric behavior, we performed first-principles calculations under finite external electric fields,[39] allowing full relaxation of both lattice parameters and atomic positions. As expected, under positive electric fields ($E > 0$), the lattice constants $a$, $b$, and $c$ all increase, indicating that LaBN$_3$ (B = Mo, W) exhibits auxetic piezoelectric effect, as shown in Fig. 3(a-b).

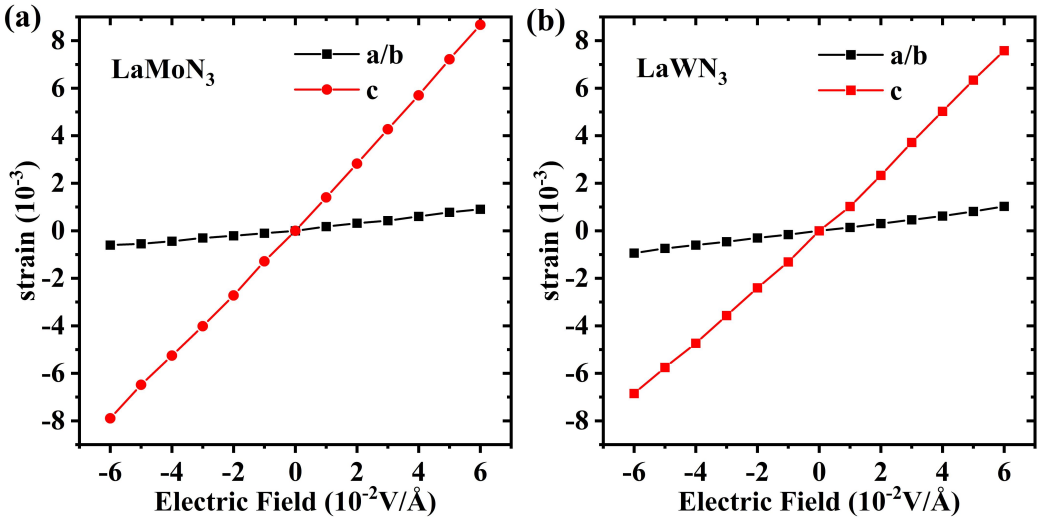

**FIG 3.** Strain in response to a finite electric field along the *c-axis* in (a) LaMoN$_3$ and (b) LaWN$_3$.

Next, we conduct an in-depth analysis of the unusual positive $e_{31}$ and $e_{32}$ coefficients. Due to the symmetry of the 3*m* point group, $e_{32}$ equals $e_{31}$. Thus, the subsequent analysis will focus on $e_{31}$. As shown in Table II, the positive $e_{31}$ value mainly originates from the dominant internal strain contribution ($e_{31}^{(i)} > 0$). To gain deeper insights into this behavior, we further decompose $e_{31}^{(i)}$ into atomic contributions, as illustrated below,[40]

$$e_{31}^{(i)} = \frac{1}{\Omega_0} \sum_k (Z_{k,31}^* \frac{du_{k,1}}{d\varepsilon_1} + Z_{k,32}^* \frac{du_{k,2}}{d\varepsilon_1} + Z_{k,33}^* \frac{du_{k,3}}{d\varepsilon_1}) \quad (1)$$

where the sum runs over atom $\kappa$ in the primitive unit cell, and $\frac{du}{d\varepsilon}$ is the displacement-response internal-strain that describes the first-order displacements resulting from a first-order strain. According to the above equation, under strain applied along the *a*-axis, the contributions of ionic displacements along the *a*- and *b*-axes to the piezoelectric response are negligible. Therefore, only the displacements of ions along the *c*-axis are considered in the following analysis. Taking LaMoN$_3$ as an example, the decomposition result of the third term is shown in Fig. 4. We observe that the Born effective charges of LaMoN$_3$ are larger than their nominal charges, indicating a obvious combination of ionic and covalent bonding characteristics, as shown in Fig. 4(a). We note that under strain applied along the *a*-axis, all cations have an upward displacement ($\frac{du_3}{d\varepsilon_1}$ (La, Mo) > 0), while N anions have a uniform direction of movement, as shown in Fig. 4(b). Based on the Born-effective charges of the cations and anions and the above equation (1), we calculated that La and Mo cations contribute +0.49 C/m² and +0.53 C/m², respectively, and N anions contribute +0.69 C/m². The total $e_{31}^{(i)}$ value is calculated to be +1.71 C/m², which is very close to the value obtained from DFPT calculations (+1.87 C/m², see Table II). Additionally, the displacement of cations and anions along *c*-axis under strain applied along *a*-axis enhances the polarization as shown in Fig. 4(d). This can be explained by the positive $e_{31}^{(i)}$ value, which is derived from the relationship $e_{31}^{(i)} = \frac{\alpha P_3}{\alpha \varepsilon_1}$.[41]

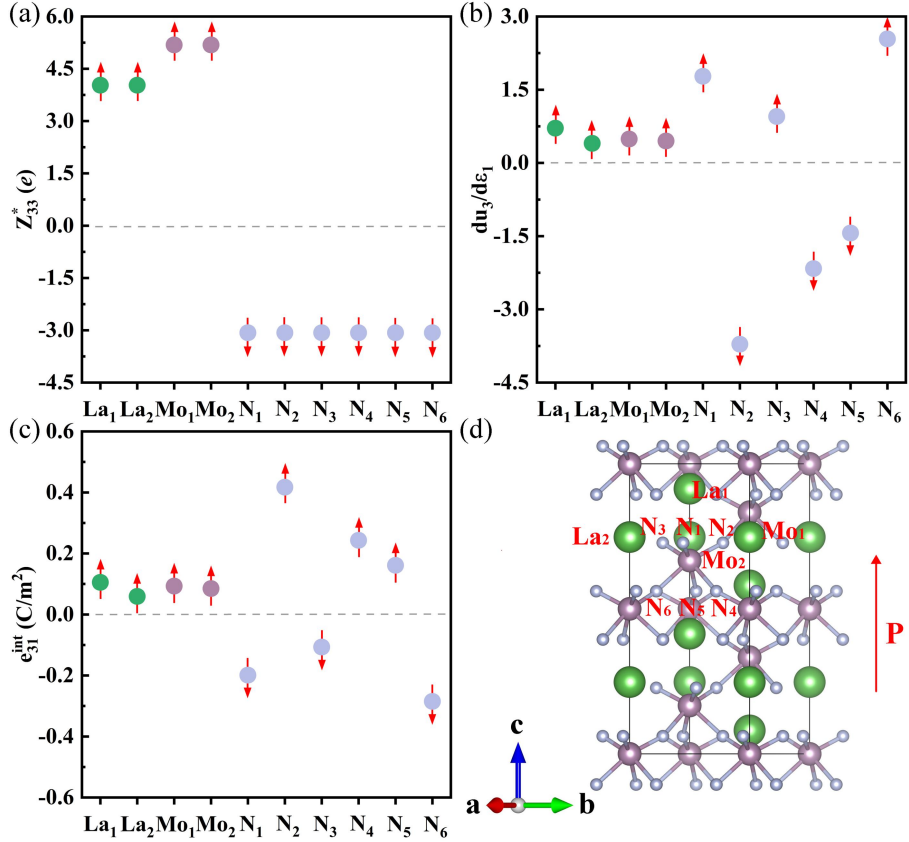

**FIG 4.** (a) Born effective charges ($Z_{33}^*$), (b) the atomic strain sensitivity $\frac{du_3}{d\varepsilon_1}$, (c) internal-strain $e_{31}^{(i)}$ and (d) the atoms of LaMoN$_3$.

**Type II Auxetic Piezoelectric Effect**. ScWN$_3$ and YBN$_3$ (B = Mo, W) exhibit unusual positive transverse piezoelectric coefficients ($e_{32} > 0$ and $d_{32} > 0$). Although YBN$_3$ (B = Mo, W) shows positive $e_{31}$ values, their corresponding $d_{31}$ coefficients remain negative (see Table II and III). In contrast, both $e_{31}$ and $d_{31}$ are negative in ScWN$_3$. Combined with the presence of positive $e_{33}$ coefficients, these features indicate that ScWN$_3$ and YBN$_3$ (B = Mo, W) possess a type II auxetic piezoelectric effect. To further confirm the presence of the auxetic piezoelectric effect, we performed first-principles calculations under finite electric fields. As illustrated in Fig. 5(a-c), when a positive electric field ($E > 0$) is applied along the polarization direction, the structures exhibit positive strain along the $b$- and $c$-axes and negative strain along the *a*-axis. This indicates that, under the influence of the electric field, one transverse direction and the longitudinal direction undergo the same trend of expansion (or contraction), while the other transverse direction exhibits an opposite deformation behavior.

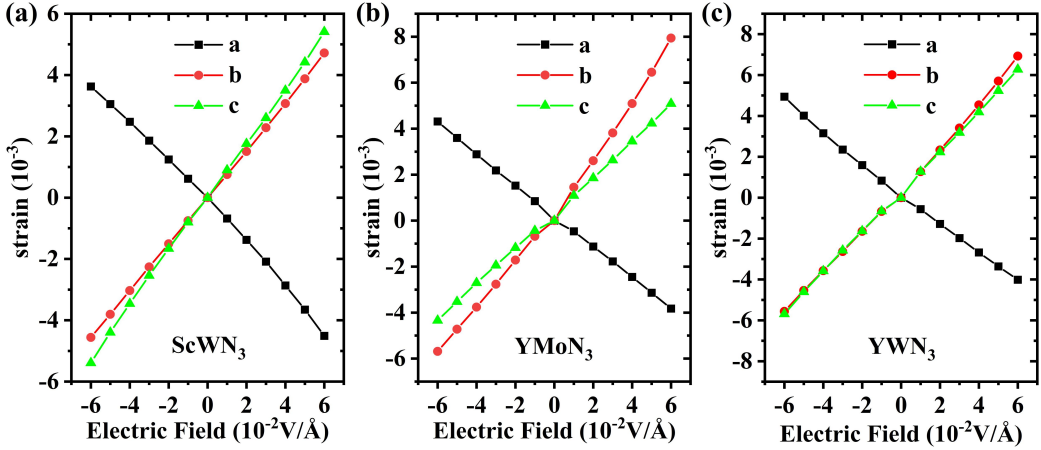

**FIG 5.** Strain in response to a finite electric field along the *c*-axis in (a) ScWN$_3$, (b) YMoN$_3$ and (c) YWN$_3$.

In ScWN$_3$ and YBN$_3$ (B = Mo, W), the positive $e_{32}$ values are primarily determined by the dominant internal-strain contribution ($e_{32}^{(i)} > 0$, see Table II). Following the same approach as before, we further decompose $e_{32}^{(i)}$ into atomic contributions, as expressed below:

$$e_{32}^{(i)} = \frac{1}{\Omega_0}\sum_k (Z_{k,31}^* \frac{du_{k,1}}{d\varepsilon_2} + Z_{k,32}^* \frac{du_{k,2}}{d\varepsilon_2} + Z_{k,33}^* \frac{du_{k,3}}{d\varepsilon_2}) \qquad (2)$$

Here, YMoN$_3$ is taken as a representative example. In the above equation, the first term sums to 0, and the contribution of the second term (0.87 C/m$^2$) is obviously lower than the contribution of the third term (1.38 C/m$^2$), so we mainly discuss the third term. And the decomposition result of the third term is presented in Fig. 6. Our results show that YMoN$_3$ exhibits positive displacement of cations ($\frac{du_3}{d\varepsilon_2}$ (Y, Mo) > 0) to strain applied along the *b*-axis, while almost all N ions except N$_1$ ions exhibit negative displacement ($\frac{du_3}{d\varepsilon_2}$ (N) < 0). Specifically, this indicates that under strain applied along *b*-axis, cations move upward, while almost all anions move downward. Based on the Born-effective charges of the cations and anions shown in Fig. 6(a) and using Equation (2) above, we calculated that Y and Mo cations contribute +0.44 C/m² and +0.42 C/m², respectively, while N anions contribute +0.51 C/m². The total contribution of these ions is positive. Additionally, the displacement of cations and anions along *c*-axis under strain applied along *b*-axis enhances the polarization as shown in Fig. 6(d). This can be explained by the positive $e_{32}^{(i)}$ value, which is derived from the relationship $e_{32}^{(i)} = \frac{\alpha P_3}{\alpha \varepsilon_2}$.

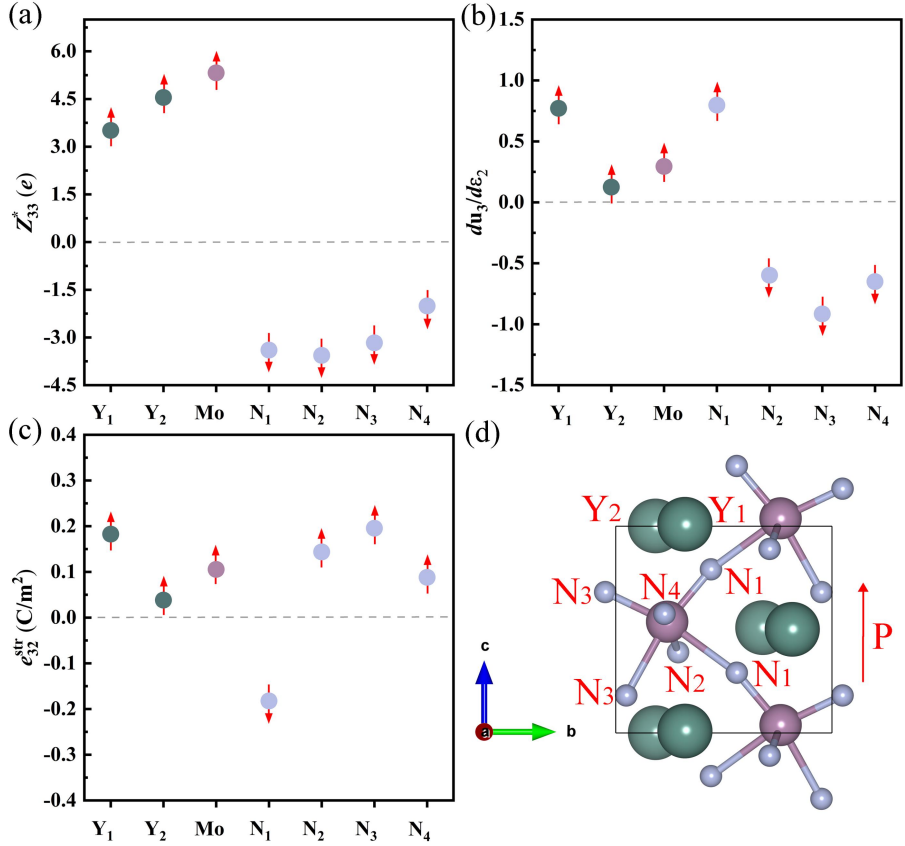

**FIG 6.** (a) Born effective charges ($Z_{33}^*$), (b) the atomic strain sensitivity $\frac{du_3}{d\varepsilon_2}$, (c) internal-strain $e_{32}^{(i)}$ and (d) the atoms of YMoN$_3$.

## IV. Conclusion

In summary, we have systematically investigated the piezoelectric properties of six nitride perovskites, specifically LaWN$_3$ and LaMoN$_3$ in the *R3c* phase, ScWN$_3$, YMoN$_3$, and YWN$_3$ in the *Pmn2$_1$* phase, and ScMoN$_3$ in the *Pna2$_1$* phase, using first-principles calculations. Among these, all compounds except ScMoN$_3$ exhibit the auxetic piezoelectric effect, which is characterized by an unusually positive transverse piezoelectric coefficient, along with a positive longitudinal piezoelectric coefficient. This behavior is in sharp contrast to previously reported results in HfO$_2$, where both the longitudinal and transverse piezoelectric coefficients are negative. These unusual positive transverse piezoelectric coefficients originate from the domination of the positive internal-strain contribution. Applying a finite electric field along (against) the spontaneous polarization direction of perovskites results in simultaneous contraction (expansion) of the lattice in both the transverse and longitudinal directions, confirming the auxetic piezoelectric effect. Our study deepens the comprehension of the piezoelectric characteristics of nitride perovskites and offers a new compositional space for the development of novel auxetic piezoelectric materials.

**Acknowledgements:**

This work was supported by the National Key Research and Development Program of China (Grant No. 2021YFA1400300), the National Natural Science Foundation of China (Grant No. 12172047), and Beijing National Laboratory for Condensed Matter Physics (2023BNLCMPKF003).